\documentstyle[11pt]{article}
\begin{document}

\def\msbar{{\rm \overline{MS\kern-0.14em}\kern0.14em}}

\begin{flushright}
 HEP-LAT Preprint: hep-lat/9209007
 Liverpool Preprint: LTH 284\\
 Edinburgh Preprint: 92/511\\
August 27, 1992
\end{flushright}
\vspace{5mm}
\begin{center}

{\LARGE\bf SU(2) Potentials from large Lattices}\\[1cm]

{\large\it UKQCD Collaboration}\\[3mm]
{\bf S.P.~Booth}\\
Department of Physics, The University of Edinburgh, Edinburgh EH9~3JZ,
Scotland

{\bf A.~Hulsebos, A.C.~Irving, A.~McKerrell, C.~Michael,
P.S.~Spencer, P.W.~Stephenson}
DAMTP, University of Liverpool, Liverpool L69~3BX, UK
\end{center}

\begin{abstract}
We measure accurate values of the inter-quark potentials on a $
48^{3}56$ lattice with SU(2) pure  gauge  theory  at $ \beta =2.85$.
The scale is set by extracting the string tension -
we obtain ${\sqrt K}a=0.063(3)$ at $\beta =2.85.$
{}From a careful study of the small-$R$ potentials in
the region 2 GeV $< R^{-1} < 5$ GeV, we extract a running
coupling constant and estimate the scale  $\Lambda _{\msbar} =
 272(24)$ MeV.
\end{abstract}

\section{Introduction}\label{Intro}
\par
The only efficient non-perturbative technique  at  present  for
theories such as QCD is lattice  simulation.   It  is  important  to
demonstrate that the limit as the lattice spacing goes  to  zero  is
precisely  understood.   Moreover,  the  physical  volume  simulated
should be kept constant as this spacing is  reduced  to
disentangle scaling violation from
finite size effects.  This programme inevitably implies
the study of lattices with many degrees of freedom and this is  only
feasible with recent increases in computing power.
\par
To study this continuum  limit  in  a  non-abelian  theory,  we
choose SU(2) pure gauge so that we can use  the  largest  number  of
lattice sites.  This work is a continuation of that reported
{}~\cite{ukqcd}  previously which was carried out using a
Meiko Computing Surface composed
of an array of 64 $i860$ processors  with
16 Mbytes of memory per processor.
The results presented  here  are  from  our final
analysis  which  yields  more precise  values   for   the
interquark potential.

 We  also  studied  glueball  and  torelon
spectra and our results (with relatively large  errors)  are
also reported here.
\par
The questions addressed in this paper are:
\begin{itemize}
  \item  Scaling: whether ratios of physical
quantities are found to be independent of the lattice spacing $a$.
  \item  The link to perturbation theory: whether lattice observables
can be accurately matched to perturbation theory calculations so enabling
the scale $ \Lambda_{\msbar} $ to be determined in terms of
non-perturbative quantities such as the string tension $K$.
\end{itemize}

\par
\section{Updating SU(2) gauge fields}
\par
We use the usual local algorithms  for  simulating  pure SU(2)
lattice gauge theory on a hypercubic lattice with periodic  boundary
conditions.  The largest size of lattice that will fit in memory,
while allowing for efficient inter-processor communications, is
$48^{3} 56.$  We estimate that we will be in the large volume region  for
such a lattice at $\beta $ values up to $\beta =2.85,$ so we choose  this
value
to be closest to the continuum limit.  Previous  work~\cite{deckandfor}
has  shown
the virtues of using an over-relaxation algorithm  (with $\omega =2)$  to
reduce the correlation between successive configurations.   This  is
very efficient to implement in the case of SU(2) and we are able  to
obtain an update time of  458~nsec  per  link.   For  the  heat-bath
algorithm, we implement a variant of the Kennedy-Pendleton method
{}~\cite{kenandpen}
using a gaussian number generator based on Ahrens and  Dieter's FL5
algorithm~\cite{ahranddiet}.
By coding in $i860$ assembler, we obtain an update time
per link of 635~nsec  where  each  link  is  updated  (so  any  link
unaccepted is retried until accepted).  Note that these update times
are about $7$ times faster than the fastest so far reported which came
from~\cite{deckandfor}
using  a  very  efficient  program  on  a   CRAY-2.
\par
We equilibrate our lattice configuration at $\beta =2.85$ with  1200
heat-bath sweeps.  We then use a ratio of  4  over-relaxation  to  1
heat-bath sweep with configurations saved for measurement every  300
sweeps.   Previous  results~\cite{deckandfor}   at $\beta =2.7$
and $2.9$  yield
autocorrelation lengths of less than 300 sweeps  for  this  updating
procedure so  we  expect  our  configurations  to  be  statistically
independent.   Subsequent  analysis   of   our   results   for   the
potential $V(R)$ at large $R$  and for the glueball correlations
confirms  that the autocorrelation time is less than 450 sweeps for all
 observables.
Our present results derive from $199$ configurations - where several
independent starts had to be made (in each case with 1200 heatbath
equilibration sweeps).
We evaluated the average plaquette action $S$ on our equilibrated
configurations and obtain $S=0.70571(5)$.
The  potential and glueball measurements  were
computed on a CRAY XMP416 at RAL.
\par
\medskip
\section{Static quark - antiquark potentials}\label{potentials}
\par
The potential between static colour sources at separation $R$ can
be obtained by measuring Wilson loops of size $R\times T$. Several
methods are known to improve the accuracy of extraction of the
ground state potential from this information. Thus a variational
path basis allows excited states to be separated more easily, a
fuzzing or blocking algorithm enhances the ground state signal, and
$T$-extrapolation techniques of varying sophistication can be used.
 \subsection{Variational path bases}

As just mentioned,  it  is
useful to consider generalised correlations of size $R\times T$  where  the
spatial paths are any gauge invariant path with cylindrical symmetry
$(A_{1g})$- not just the straight line used in a rectangular Wilson loop.
Moreover  these  two  spatial  paths,  at  times $t=Ta$  apart,  may  be
different.  Hence the measurable will be $C_{ij}(R,T)$ where $i,j$ label the
path type.  Then the largest eigenvalue $\lambda (R)$ of the transfer  matrix
in the presence of the static sources is related to the potential by
\par
$$
aV(R) = - \log ( \lambda (R) )
$$
\noindent and $\lambda (R)$ can be obtained from considering the limit
 as $T\to\infty $ of  the
largest eigenvalue $\lambda (R,T)$ of the equation
\par
$$
C_{ij}(R,T) u_{j} = \lambda (R,T) C_{ij}(R,T-1) u_{j}
$$
\noindent This is a variational method where an optimum combination of
paths $i$
(with $i=1,..N)$ is chosen. At larger $T$, statistical errors may cause
the eigenvalue analysis to become unstable. Thus a more robust procedure
is to extract the optimum path basis $u_i$ from modest $T$-values such
as $2/1$ and then extract the appropriate path combination ${\cal C}$
for study of larger $T$:
$$
 {\cal C}(R,T)= u_i C_{ij}(R,T) u_j
$$
 and effective eigenvalue and potential values defined by
$$
 \lambda_{eff}(R,T)={{\cal C}(R,T) \over {\cal C}(R,T-1)}= \exp
(-V_{eff}(R,T)a)
$$
 \noindent The best  choice  of  paths  would  allow
$\lambda_{eff} (R,T)$ to be close to its asymptotic value for small T.
  In this way
the statistical errors  will  be  smallest.
\subsection{Fuzzing}

  As a measure of the efficiency of a path basis in extracting
the ground state, it  is  convenient  to
introduce an \lq\lq overlap" defined as
\par
$$
{\cal O}(R) = \prod _{T}\{ \lambda_{eff} (R,T)/\lambda (R) \}.
$$
\noindent For straight paths of length $R/a=24$ , we find ${\cal
O}<0.05.$
   Thus
rectangular Wilson loops are a very inaccurate way of extracting the
ground state potential.
As has been known for some time, a valuable improvement comes
from fuzzing or blocking these
paths~\cite{teper,albanese,perandmich91}.
Using a purely spatial
blocking allows the transfer matrix interpretation to  be  retained.
In order to reduce  the  memory  and  storage  requirements  of  our
lattice configurations, we performed one  step  of  Teper
blocking~\cite{teper}
which gives an effective spatial lattice of size $24^{3}$.   We  did  not
vary the parameters of this initial blocking step, but  chose  equal
weights for the 5 paths (double straight  path  and $4 U$-bends)  as
found to be satisfactory
previously~\cite{teper,perandmich91}.  Because of  this  initial
data smoothing we can only evaluate  potentials  at  separations  an
even distance apart.  For further smearing we used a  more  flexible
scheme with no factor of two  scale  change:  namely  the  recursive
blocking scheme~\cite{albanese}:
\par
$$
U(new) = {\cal P}_{SU(2)} [ c~U(\hbox{straight}) +
\sum^{4}_{1}U(\hbox{u-bends})]
$$
\noindent After exploratory studies to optimise the overlap ${\cal O}$,
we chose $c=2$
and a maximum of $110$ iterations of this recursive blocking.   Larger
$c$ values gave slightly bigger overlaps but at the  expense  of  much
larger iterations (eg. 150 iterations at $c=4)$.  As  a  variational
basis with $N=3,$ we use 110, 70 and 30 iterations  for  the  paths.
This basis for $R/a = 24$ yields an overlap ${\cal O}=0.92$ which is  a  huge
improvement over the purely unblocked case.  Because this  value  of
the overlap is close to $1.0$, one can get reliable estimates  of  the
ground state potential from  relatively small $T$-values which have
smaller errors.
\subsection{$T$-extrapolation}
\par
In order to extract the ground state potential values, one has to
extrapolate the effective eigenvalues $\lambda(R,T)$ to large $T$.
This will be controlled by the eigenstates of the lattice transfer
matrix with
$$
C_{ij}(R,T)=b^0_i b^0_j \lambda_0^T (R) + b^1_i b^1_j \lambda_1^T (R) +..
$$
where for a given $R$-value, $\lambda_0$ is related to
the ground state potential as $\exp(-Va)$
and  $\lambda_1 / \lambda_0 = \exp(-\Delta Va)$ with $\Delta V$ the energy
gap to the first excited state.
Then the extrapolation of the effective eigenvalue will be controlled
by the admixture of excited
state contributions, since
$$
\lambda_{eff}(T)= { \lambda_0^{T+1} + b \lambda_1^{T+1} + .. \over \lambda_0^T
+ b \lambda_1^T + ..}
  =  \lambda_0 \left( 1-b {\left( \lambda_1 \over \lambda_0 \right)}^T
  \left( 1- {\lambda_1 \over \lambda_0}+ ...\right)  \right)
$$
Since we use the same path operator (the linear combination
corresponding to the 2/1 eigenvector with largest eigenvalue)
at each end, we will have $b > 0$ which ensures that $\lambda_{eff}(T)$
monotonically increases to the asymptotic value $\lambda_0$ or
equivalently that $V_{eff}(R,T)$ monotonically decreases to the asymptotic
 value $V(R)$. Now the appropriate
energy gap is that to the first excited state that couples to the operator
used for the path of length $R$. To avoid contamination by other
path representations of lower energy, one should  use
the most symmetric path
operators (with an $A_{1g}$ representation under the group $D_{4h}$ ), then
the energy gap  $\Delta V$ will be given by the next such $A_{1g}$ state.
 Previous  work~\cite{perhuntandmich,perandmich91} has already
determined the energy difference $\Delta V$
between the first excited $A_{1g}$ state and  the  ground  state  we  are
trying to isolate.  Moreover this determination agrees with the string
model value of $2\pi/R$ which should be an appropriate estimate at
large $R$. This implies that the problem of excited state contamination
in the signal will be most important at large $R$ where the energy gap
is least.
However, even   at  large $R$, $\Delta Va\approx 0.2$ and then any substantial
  contamination  of  this  excited
state would be very easily detectable as a steady  decrease  of  the
effective potential value with increasing $T$.
\subsection{Data analysis}
\par
We have results from measurements of 199 configurations. We combine
these into 39 blocks of 5 adjacent measurements. Each such a block covers
1500 update sweeps and we find no statistically significant evidence for
any autocorrelation between blocks. There is a strong correlation
between the measurements of different sized Wilson loops, however. We
take this into account primarily by using a bootstrap analysis of
statistical errors. Thus we choose 39 blocks randomly (with
substitution) from the 39 available and for each such choice we
repeat the analysis of any quantity of interest. The variance over
many such random selections then gives the error distribution
required.

To estimate the potential $aV(R)$, we proceed as outlined above.
With our variational basis of 3 paths (from the 3 different fuzzing
levels), we can estimate the ground state and first excited state
eigenvalues and eigenvectors of the transfer matrix even at relatively
low $T$-ratios such as $2/1$. Since the statistical error is here
quite small, we use this determination of the eigenvectors to
fix the linear combination of the paths to be used at larger $T$.
We then obtain effective potential values for each adjacent $T$-ratio
and these are given in Table~\ref{tablepot}.  To extrapolate to
 large $T$, we use
two different methods. As a first method we identify the start of the
plateau: namely the $T$-value at which the $(T+1)/T$ and $T/(T-1)$
effective potential values are equal within errors. Because of the
strong correlation of the errors on these two quantities, we perform
a bootstrap error analysis of the difference to ascertain the start of
the plateau when this difference is statistically consistent with zero.
In the preliminary analysis of part of this data \cite{ukqcd},
we found that the $4/3$ $T$-ratio was sufficient to give the plateau
value.  Because of the larger statistics in this work,
we find somewhat larger $T$-values are needed in some cases.
Indeed we find that the effective potential from $5/4$ is always
consistent with being  equal within one standard deviation to the
$6/5$ effective potential. Thus the
plateau value can be identified with the $5/4$ $T$-ratio. Note that
since the effective potential is monotonically decreasing with $T$, the
$5/4$ estimate is strictly an upper bound.

Because the excited state contamination is expected to be relatively
larger at larger $R$, as discussed above, we may expect that the
potential is overestimated more at large $R$ so yielding a larger
string tension value. In order to account for this,
the other method we use is an attempt to extrapolate to $T = \infty$
using the information on the energy gap to the first excited state to
stabilise this extrapolation.  We are able to control this extrapolation
because we have an estimate of the energy gap to the first excited
state and we know that the effective potential has a monotonic
decrease with increasing $T$. Thus by assuming
that the approach to asymptopia is as slow
as possible (ie it is given entirely given by the lowest excited state),
we can extract the maximum systematic error in the evaluation of
the ground state potential. This is most probably an overestimate
of the contribution of the first excited state since the variational
analysis selects a path combination which is designed to separate
out the contributions from the ground state and first excited state -
and if this has been accomplished then $b=0$.

  Fitting the effective potential values
for $T$ from 2 to 7, we find an acceptable fit (using the full
covariance matrix) with the excited state energy gap fixed from the
$2/1$ variational analysis. In fact this fit gives very similar results
to solving exactly for the two parameters $\lambda_0$ and $b$
using the effective potentials from $3/2$ and $4/3$ only. These results
are shown in table~\ref{tablepot}. One indeed sees evidence
 for a sizeable systematic error particularly at larger $R$.
  One feature worth noting
is that the difference $dV$ between the plateau value $V(R,5)$ and the
extrapolated value  $V_{\infty}(R)$  is itself not smooth with $R$.
It should be possible to use the reasonable requirement of
continuity versus $R$ to constrain the correction somewhat - we have
not pursued this.

\begin{table}
\centering
\begin{tabular}{|c|c|c|c|c|c|c|}\hline
 $R/a$  &   $aV(R,3)$  & $aV(R,4)$ & $aV(R,5)$ &$aV(R,6)$ &$aV(R,7)$ &
$aV_{\infty}(R)$  \\\hline
 $2$ & 0.3542(1) & 0.3540(1) & 0.3538(1)& 0.3538(2\ )&0.3536(3)&0.3534(2\ )\\
 $4$ & 0.4181(1) & 0.4178(1) & 0.4176(2)& 0.4179(2\ )&0.4181(4)&0.4169(4\ )\\
 $6$ & 0.4461(1) & 0.4456(2) & 0.4454(3)& 0.4456(4\ )&0.4458(5)&0.4442(4\ )\\
 $8$ & 0.4645(2) & 0.4639(2) & 0.4639(2)& 0.4640(3\ )&0.4634(5)&0.4619(5\ )\\
$10$ & 0.4791(2) & 0.4782(3) & 0.4780(3)& 0.4783(5\ )&0.4782(7)&0.4754(7\ )\\
$12$ & 0.4915(3) & 0.4905(4) & 0.4906(3)& 0.4909(5\ )&0.4900(6)&0.4866(8\ )\\
$14$ & 0.5028(4) & 0.5017(4) & 0.5009(4)& 0.5009(6\ )&0.5013(9)&0.4976(9\ )\\
$16$ & 0.5134(5) & 0.5122(5) & 0.5120(5)& 0.5123(6\ )&0.5118(10)&0.5076(10)\\
$18$ & 0.5234(5) & 0.5220(6) & 0.5211(6)& 0.5218(7\ )&0.5196(10)&0.5165(11)\\
$20$ & 0.5333(7) & 0.5314(7) & 0.5301(8)& 0.5302(10)&0.5296(11)&0.5239(14)\\
$22$ & 0.5428(8) & 0.5409(8) & 0.5399(9)& 0.5397(10)&0.5361(11)&0.5328(18)\\
$24$ & 0.5522(9) & 0.5502(9) & 0.5489(9)& 0.5496(11)&0.5481(13)&0.5415(20)
 \\\hline
\end{tabular}
\caption{ The effective potentials at $\beta=2.85$ between static sources at
separation $R$. Here $V(R,T)$ is determined from the $T/(T-1) $ ratio
of correlations as described in the text.
 $V_{\infty}(R)$ is determined from extrapolating the 3/2
and 4/3 effective potential values as described in the text. }
\label{tablepot}
\end{table}

\subsection{Off-axis potentials}
 \par
As well as the preceding high statistics method of obtaining the potential
between sites an even number of lattice spacings apart (and along
a lattice axis), we carried out a complementary study of potentials
at off-axis separations.  We choose vector separations $(R_x,R_y,R_z)$
with $0 \le R_i \le 3a$. For the 2 and 3 dimensional paths,
we sum over all 2 (6) symmetric routes along the edges of the
2 (3) dimensional hypercuboid. We used the full $48^3$ spatial
configuration and constructed operators by repeated blocking of links with
{}~\cite{albanese} $c=2$ and $100$ iterations. This level of blocking
is less than that described above but it is optimal for the
relatively shorter paths considered here. Indeed at $R=2a$ where a
direct comparison can be made, we find overlaps of 99.5\% here whereas
only 97.9\% with the previous fuzzing scheme. Because of the computational
and data communication constraints, we carried out this small-$R$
measurement for 8 configurations only. Nevertheless, the statistical
errors are very small. We processed these results using the method of
identifying a plateau as described above and we found that the
4/3 and 5/4 $T$-ratio effective potentials were equal within errors in
every case. These results are shown in table~\ref{table3d}.
 In subsequent analyses, we used the 4/3 $T$-ratio values.

\begin{table}
\begin{center}
\begin{tabular}{|c|c|c|c|c|}\hline
 $R_x/a$  & $ R_y/a$ & $R_z/a$  & $aV(R,4)$ & $aV(R,5)$ \\\hline
1 & 0 & 0  & 0.2597(1\ )   & 0.2596(1\ ) \\
1 & 1 & 0  & 0.3226(1\ )   & 0.3225(1\ ) \\
1 & 1 & 1  & 0.3489(2\ )   & 0.3489(3\ ) \\
2 & 0 & 0  & 0.3538(2\ )   & 0.3537(2\ ) \\
2 & 1 & 0  & 0.3699(2\ )   & 0.3697(2\ ) \\
2 & 1 & 1  & 0.3802(2\ )   & 0.3805(3\ ) \\
2 & 2 & 0  & 0.3917(3\ )   & 0.3916(4\ ) \\
2 & 2 & 1  & 0.3972(4\ )   & 0.3976(5\ ) \\
2 & 2 & 2  & 0.4083(5\ )   & 0.4081(6\ ) \\
3 & 0 & 0  & 0.3944(4\ )   & 0.3942(5\ ) \\
3 & 1 & 0  & 0.3998(4\ )   & 0.3998(5\ ) \\
3 & 1 & 1  & 0.4040(4\ )   & 0.4043(4\ ) \\
3 & 2 & 0  & 0.4105(5\ )   & 0.4107(6\ ) \\
3 & 2 & 1  & 0.4134(4\ )   & 0.4137(6\ ) \\
3 & 2 & 2  & 0.4207(5\ )   & 0.4209(6\ ) \\
3 & 3 & 0  & 0.4224(5\ )   & 0.4221(7\ ) \\
3 & 3 & 1  & 0.4246(5\ )   & 0.4246(6\ ) \\
3 & 3 & 2  & 0.4294(6\ )   & 0.4295(8\ ) \\
3 & 3 & 3  & 0.4358(7\ )   & 0.4356(9\ )  \\\hline
\end{tabular}
\caption{ The potentials at $\beta=2.85$ between static sources at
separation $(R_x,R_y,R_z)/a$. Here $V(R,T)$ is determined
 from the $T/(T-1) $ ratio of correlations.}
\label{table3d}
\end{center}
\end{table}

\subsection{$R$-dependent fits}
\par
We concentrate first on extracting a string tension value from our
high statistics data.  To avoid the influence of the form of the fit at
smaller $R$, we    fit  the  static  potential  for $R \ge 6a$   with   the
conventional lattice-Coulomb plus linear term:
\par
$$
V(R) = C - {E\over R} + K R
$$
\noindent A lattice-Coulomb expression for the lattice one gluon
propagator on an infinite lattice is used for $1/R$ although
in practice this is very close to $1/R$ for $R \ge 6a$. Since the potential
values at nearby $R$-values are highly correlated, we take into account
the full covariance matrix in minimising $\chi ^2$.
Using a fit to the previously determined values of $V(R)$  and  taking
into account correlations in errors between potentials at  different
$R$-values, we find the result of table~\ref{params}.
The fit is illustrated in fig~\ref{fforce}.   The errors on the
potential values are somewhat less correlated when expressed as
difference values related to the force $(V(R+a)-V(R-a))/2a$  and these
data are collected in table~\ref{tablefor} to allow other fits to
 be considered.

\begin{table}
\begin{center}
\begin{tabular}{|c|c|c|c|}\hline
   &   $E$      &      $Ka^2$ & $\chi^2 / {\rm d.o.f.}$\\\hline
 $V(R,5)$  &  0.247(7)  &     0.00401(8)&   15.2/7 \\
 $V_{\infty}(R)$ & 0.247(14) & 0.00363(17)&  8.2/7  \\\hline
\end{tabular}
\caption{The fitted parameters to the potential for $R\ge 6a$.}
\label{params}
\end{center}
\end{table}

\begin{table}
\begin{center}
\begin{tabular}{|c|c|c|}\hline
 $R/a$   & $aV(R+2a,5)-aV(R,5)$&$aV_{\infty}(R+2a)-aV_{\infty}(R)$\\\hline
 $2$    & 0.0638(2)    & 0.0635(3\ )\\
 $4$    & 0.0278(2)    & 0.0273(5\ )\\
 $6$    & 0.0185(2)    & 0.0177(5\ )\\
 $8$    & 0.0141(2)    & 0.0134(5\ )\\
$10$    & 0.0125(2)    & 0.0112(5\ )\\
$12$    & 0.0103(2)    & 0.0110(7\ )\\
$14$    & 0.0111(4)    & 0.0100(7\ )\\
$16$    & 0.0091(4)    & 0.0088(7\ )\\
$18$    & 0.0090(4)    & 0.0074(7\ )\\
$20$    & 0.0098(4)    & 0.0089(9\ )\\
$22$    & 0.0090(4)    & 0.0089(14) \\ \hline
\end{tabular}
\caption{ The potential differences at $\beta=2.85$ between static sources at
separation $R$}
\label{tablefor}
\end{center}
\end{table}

{}From these fits, we see the expected feature that the extrapolated
 potential values
yield a smaller string tension since the extrapolation has most
impact at large $R$. A reasonable approach would be to use the
difference between the two string tension values as an estimate
of the systematic error. Indeed the $V(R,5)$ values are
upper bounds whereas the $V_{\infty}$ values are very much extreme
lower estimates. So the true value of the string tension should
lie between the values extracted from our two fits. Since the
$V(R,5)$ analysis has the smaller statistical
errors, we quote $K=0.00401(8)(38)$. This is effectively a $10 \% $
error on $K$. Note that the $\chi ^2 $ value is poor
 for the fit to the $T$-ratio 5/4 effective
potential. This is part of a general phenomenon that at larger
$T$-values the errors seem to be non-gaussian.

As well as the errors from statistics and from $T$-extrapolation, there
may also be errors due to the choice of function $V(R)$ to fit. To
be specific, we have fitted for $ 0.38 < R\sqrt K < 1.52 $ with a
simple two parameter fit to the force (potential differences). If there were
$\log(R)$ terms present as well, then the string tension would be
much harder to extract.
The fitted value for the string tension enables  us  to
set the scale by requiring  $K=0.44$ GeV, so yielding $a^{-1}=6.95(35)$
GeV.  Relating $a$ to $\Lambda_L$ by the two-loop perturbative
$\beta$-function then gives  ${\sqrt K}/\Lambda _{L}=43.0(2.1)$
at $\beta =2.85.$

We also wish to understand the potential at all $R$ and we follow
the method used in \cite{cmlms} to make a comprehensive fit to on-
and off-axis potential values.
The potential shows a lack of rotational invariance at small $R$.
To lowest order this can be attributed to the difference $\delta G(R)$ between
the lattice one gluon exchange expression and the continuum expression.

$$
\delta G(R) =  {4 \pi \over a} \int_{-\pi}^{\pi}
 { d^3 k \over(2\pi)^3}
{ e^{ik.R/a} \over 4\sum_{i=1}^3 \sin^2 (k_i / 2) }
-{1 \over R}
$$
On a lattice, the next order of perturbation has been calculated
\cite{hellandkar} for $R$ values on-axis
and the dominant effect is a change from the bare coupling
to an effective coupling \cite{lm}. In that case,  using the difference
above but with an adjustable strength will correct for the small $R/a$
lack of rotational invariance. A test of this will be that a smooth
interpolation  to the  off-axis potential value of
 $V(R)$ versus $R$ is obtained with this one free
parameter.

  We  evaluate $\delta G(R)$
numerically using the limit of a very large lattice since we are not
here concerned with long-range effects.  Then we use the  following
empirical expression to fit for $R~> a$,
$$
 V(R)=C- {A \over R}+{B \over R^2}+KR - A f \delta G(R)
$$

\noindent Here $f=1$ and $B=0$ corresponds to the previous two parameter
fit where $E=A$.  Now we fix the parameter $K$ from the foregoing
large $R$ analysis, and fit to the smaller $R$ data with the above
expression. We do not have sufficient data samples
to use the full covariance matrix, so we fit to the force between
adjacent $R$-values using the data for $V(R,4)$ of Table~\ref{table3d}.
 We find $\chi ^{2}$ per degree of
 freedom 9.7/14 which is acceptable. The fit parameters are
shown in table~\ref{par3d}. The good fit quality shows
that the lack of rotational invariance is indeed fully
accommodated by the rather simple minded model of one lattice-gluon
exchange.  This is illustrated in  fig~\ref{offax} where the ratio of fit
to data versus $R$ is shown.

\begin{table}
\begin{center}
\begin{tabular}{|c|c|c|c|}\hline
$A$ & $f$ & $Ka^2$ & $B/a$ \\\hline
 0.238(7) & 0.63(3) & 0.00401 & 0.050(6) \\\hline
\end{tabular}
\caption{Fit to force for $R>a$.}
\label{par3d}
\end{center}
\end{table}
\par
 For our present purposes
the detailed form of this fit at small $R$ is not relevant
 - what is needed is a confirmation that a good fit
can be obtained. This then supports  our prescription to
 correct the lattice artifacts responsible for the lack of rotational
invariance. What is more difficult is to assign errors to this
correction procedure. Since the correction coefficient $f$ is determined
to a few per cent, the statistical errors are small. The fact that one
parameter corrects all the off-axis points simultaneously is very
encouraging.

 So in conclusion, we find that the lattice artifact effects from
lack of rotational invariance can be well accommodated by the one
gluon exchange correction with $f=0.63(3)$ and hence the
lattice potential data can be
corrected for this effect explicitly. This conclusion agrees with
that ~\cite{cmlms} at $\beta=2.7$ where $f=0.68(3)$ was obtained.
\par
\par
\section{ Scaling}
\par
These results can be compared with a similar
analyses~\cite{perandmich91,cmlms} on $32^{4}$ lattices  at $\beta =2.7$
and on $24^4$ lattices~\cite{cmtok} at $\beta=2.4$ for which
the potentials have been calculated extensively.
A particularly explicit way to test scaling is to plot the force
versus $R$ at different $\beta$-values and match the resulting curves
to one universal curve. The advantage of using the force is that
self-energy effects cancel.
In order to expose the scaling (ie $a$-dependence), we construct
from the lattice artifact corrected potentials the
dimensionless ratio
$$
\alpha( { R_1 + R_2 \over 2 }) = { 4\over 3} R_1 R_2 { V_c(R_1)-V_c(R_2) \over
 R_1-R_2 }
$$
(the factor 4/3 is just a convention here). It
turns out that the  error in using a finite difference in $R$ is
 negligible for this
quantity in our subsequent use. The advantage of this quantity
$\alpha(R)$ is that it is easy to visualise the scaling between
different
$\beta$-values since the $R$-axis only has to be scaled.

 Now we have access to accurate data
on the force at $\beta=$2.4 from ref~\cite{cmtok} and 2.7 from ref~
\cite{cmlms}. We consider the common physical $R$-range $0.4 < R
 \sqrt K <1.6$ and determine the $R$-scaling factor at 2.4 and 2.7
needed to bring the data to overlap best the 2.85 results  from
table~\ref{tablefor}.
The comparison is shown in fig ~\ref{scaling} where
we have used the best fit ratios of lattice spacings: $a(2.4)/a(2.85) =
4.12(2)$ and $a(2.7)/a(2.85)=1.58(1)$ (these  errors do not
include the systematic error from $T$-extrapolation at 2.85).
We see that a universal curve describes the data well. This is especially
impressive given that the ratio of lattice spacings is as large as 4.
Thus scaling is well satisfied by the interquark potential for
$\beta \ge 2.4$.

A more detailed study of the ratio $a(2.7)/a(2.85)$ is possible
once scaling is accepted. The most reliable determination comes from the
force at moderate $R$: at small $R$ the lattice artifact corrections
are too big while at large $R$ the $T$-extrapolation errors are large.
We use $\alpha(R=5a(2.85))$ as the best determined point. We obtain
0.4448(32) from the 5/4 $T$-extrapolation and  0.4368(80) from $V_{\infty}$.
The lattice artifact correction (using $fA=0.150$ from our fit of
table~\ref{par3d}) is
-0.0090. Thus we determine 0.436(8) where the error covers statistical,
$T$-extrapolation and lattice artifact correction. The lattice artifact
corrected fit~\cite{cmlms} to the $\beta=2.7$ data gives an
interpolation in terms of $r=R/a(2.7)$:
$$
 \alpha(r)= 0.348 - 0.144/r +0.01373 r^2
$$
Assuming that the errors on this expression are similar (ie
2\% on $\alpha$ at $\alpha$ near 0.436), we obtain
 $r=3.12(13)$. Hence we obtain
$a(2.7)/a(2.85)=1.60(6)$. This ratio agrees with that illustrated
above but now all errors are included.

As found previously ~\cite{ukqcd}, we find that asymptotic scaling
is not valid at these $\beta$ values. Indeed since two-loop asymptotic
scaling implies \mbox{$a(2.7)/a(2.85)=1.464$}, the above $a$-ratios already
exclude asymptotic scaling. Another, more traditional, way to get at
this ratio is to compare the string tension results at the two
$\beta$-values. This has the disadvantage that an extrapolation in $R$
is required but is useful since the string tension is easy to compare
between different lattice analyses. In order to try and reduce
the systematic errors in $T$-extrapolation, we use similar analyses
at the two $\beta$-values.
  Thus fitting in the same $R$-range ( $0.4 < R
 \sqrt K <1.6$~) and (a) using $T$-ratio 5/4 we get string tension $Ka^2$=
0.0109(2) and 0.00401(8); while (b) using the extrapolated value from
4/3 and 3/2 ratios gives 0.0103(2) and 0.00363(17) respectively at
$\beta=2.7 $ and 2.85.
Thus we   evaluate  $a(2.7)/a(2.85)=1.65(5)$
 from  the  respective  string  tension  measurements.

 These discrepancies with asymptotic scaling can be expressed usefully
 in terms of the $\beta $-function.  The perturbative expression for the
effective $\beta $-function is
$$
{-{\Delta \beta} \over{ \Delta \ln a}} = {11\over{ 3\pi^2}}
[1 + {{17 g^2}\over{ 44\pi^2}} +\dots]
$$
\noindent which is $0.393$ at $\beta =2.7$ and $0.392$  at  $2.85$. Thus
an accurate determination of $a(2.7)/a(2.85)$ will give the required
$\beta$-function.
Using $a(2.7)/a(2.85)=1.60(6)$ gives $-\Delta \beta /\Delta \ln a=0.319(25)$
 which  is  only $81\% $  of  the  two  loop
perturbative result.

  Thus we conclude that  asymptotic  scaling  of
the string tension  is  not  found  up  to $\beta =2.85.$   Moreover  this
effective $\beta $-function from 2.7 to 2.85 is rather close to that  found
at larger lattice spacing (e.g. from a comparison of $\beta =2.5$ and $2.7:
-\Delta \beta /\Delta \ln a=0.318(10)$  from  a  string  tension
determination~\cite{perandmich91,cmlms} and
0.310(6) from a MCRG analysis~\cite{deckandfor}).
\par

\section{Glueballs and Torelons}
\par
We also have measured glueball and torelon correlations on our 199
configurations. The operators were those used by ref~\cite{perandmich91}
at $\beta=2.7$ except that another iteration of fuzzing was employed.
The error analysis comes from a bootstrap evaluation of 39 blocks of
5 configurations each (as in the potential case). The effective masses
in the 1/0 variational path basis are shown in table~\ref{tablegb}. We
see restoration of rotational symmetry (ie $E^+$ and $T^+_2$ representations
of $O_h$ are degenerate and form the 5-dimensional $J^P=2^+$ state, etc). It
is reasonable to take the 3/2 $T$-ratio effective masses as an estimate
with a systematic error given by the differences between the results
from different $T$-ratios.
To compare with the results at $\beta=2.7$, for the 3/2 effective mass,
we quote the ratio
 $a(2.7)/a(2.85)$ obtained: namely 1.70(16) for $0^+$, 1.57(8) for $2^+$ and
1.68(9) for A(100) torelon. These ratios are consistent with each other
and with the scaling found from the inter-quark potential in the
previous section.
 For the heavier states, we also confirm the results found at $\beta=
2.7$, in particular the signal for a $J^P=1^-$ state (from the $T^-_1$
representation) which has exotic spin-parity.

{}From the $A(100)$ torelon energy, we can find the effective string
tension $K_{48}$ of the colour flux loop which encircles
 the spatial boundary
of length $L=48a$. This is $K_{48}a^2=0.00360(15)$. Correcting for
string fluctuation effects then allows an estimate of the string
tension $K=K_{48}+\pi /(3 L^2) =0.00405(15)a^{-2}$. This latter value is
in excellent agreement with our result from analysis of the heavy
quark potential.

\begin{table}
\begin{center}
\begin{tabular}{|c|l|l|l|l|l|}\hline
 Rep  &   1/0  & 2/1 & 3/2 & 4/3  & 5/4   \\\hline
 $A^+_1$  & 0.336(11)   & 0.252(9)    & 0.229(12)  & 0.194(17) &0.196(18) \\
 $E^+  $  & 0.529(11)   & 0.377(12)   & 0.357(16)  & 0.358(27) &0.373(13) \\
 $T^+_2$  & 0.533(8)   & 0.393(12)   & 0.350(16)  & 0.311(19) &0.343(24)\\
 $A^-_1$  & 0.690(19)   & 0.430(26)   & 0.448(35)  & 0.403(57) &0.286(73)\\
 $E^-  $  & 0.744(13)   & 0.505(16)   & 0.465(30)  & 0.428(35) &0.340(65)\\
 $T^-_2$  & 0.752(11)   & 0.533(14)   & 0.510(25)  & 0.505(38) &0.422(83)\\
 $A^+_2$  & 0.834(18)   & 0.588(36)   & 0.620(53)  & 0.687(144)&         \\
 $T^+_1$  & 0.976(15)   & 0.626(27)   & 0.591(57)  & 0.534(71) &         \\
 $A^-_2$  & 1.224(41)   & 0.713(47)   & 0.602(108) & 0.649(183)&         \\
 $T^-_1$  & 0.993(18)   & 0.677(21)   & 0.654(54)  & 0.640(88) &0.694(167)\\
 A(100) & 0.248(7)      & 0.176(6)    & 0.173(7)   & 0.167(8)  &0.155(8) \\
 A(110) & 0.452(16)     & 0.301(19)   & 0.268(21)  & 0.267(21) &0.243(31)\\
 A(111) & 0.696(69)     & 0.355(55)   & 0.292(53)  & 0.268(82) &0.275(70)\\
\hline
\end{tabular}
\caption{ The effective glueball and torelon masses  at $\beta=2.85$
 determined from the $T/(T-1) $ ratio
of correlations as described in the text.
}
\label{tablegb}
\end{center}
\end{table}

\section{Running coupling}
\par

In the continuum the potential between static quarks is known
perturbatively to two loops in terms of  the  scale $\Lambda_{\msbar} $.
 For  $SU(2)$ colour, the continuum force is given by
$$
{dV \over dR } =  {3 \over 4} {\alpha(R) \over R^2}
$$
\noindent with the effective coupling $\alpha (R)$ defined as
$$
\alpha (R) = { 1 \over 4 \pi [ b_0  \log (R\Lambda _R )^{-2} +
(b_1 / b_0 ) \log \log (R\Lambda_R )^{-2} ] }
$$
\noindent where $b_0=11/24 \pi ^2$ and $b_1=102 \ b_0^2/121$ are
 the usual coefficients in the
perturbative expression for the $\beta$-function
and, neglecting quark loops in the vacuum,  $\Lambda_R= 1.048
 \Lambda_{\msbar}$ (note that there is an error in this numerical
value as quoted in ref.\cite{cmlms}).
Note that the usual lattice regularisation scale $\Lambda_L = 0.05045
 \Lambda_{\msbar}$.
\par
It is now  straightforward  to  extract  the  running  coupling
constant from our lattice potentials by using~\cite{cmlms}
$$
\alpha( { R_1 + R_2 \over 2 }) = { 4\over 3} R_1 R_2 { V_c(R_1)-V_c(R_2) \over
 R_1-R_2 }
$$
\noindent where the error in using a finite difference is here negligible.
Here $V_c(R)$ are the potentials corrected for lattice artifacts
as discussed above. We assign a systematic error of $10\%$ to this
correction as an illustration.
These values are shown in table~\ref{alpha}
 and are plotted  in fig~\ref{falpha} versus
$R \sqrt K$ where $K$ is taken from the fit - see table~\ref{params}.
The interpretation of $\alpha$ as defined above as an
effective running coupling constant is only justified at small $R$ where the
perturbative expression dominates.  Also shown  are  the
two-loop perturbative results for $\alpha(R)$ for
different values of $\Lambda_R $.

The figure clearly shows a {\it running} coupling constant.  Moreover
the result is consistent with the expected perturbative dependence  on $R$  at
small $R$.  There are systematic errors, however. At larger $R$, the
perturbative two-loop expression will not be an accurate estimate
of the measured potentials, while at smaller $R$, the lattice artifact
corrections are relatively big.  Setting the scale using $\sqrt K=0.44$ GeV
implies $1/a(2.85)=7.0 $ GeV, so $R < 3.5a(2.85)$ corresponds to
values of $1/R > 2$ GeV.  This $R$-region is expected to be adequately
 described
by perturbation theory.  Another indication that perturbation
theory is accurate at such $R$-values is that $\Delta V_c / \Delta R$
at small $R$ is found to be very much greater than the non-perturbative
 value $K$ at large $R$.

The curves shown in fig~\ref{falpha} allow an estimate of the running
coupling at small $R$. In practice, $\alpha(R)$ is numerically
nearly equal to $\alpha_{\msbar}$ at a momentum scale $q=1/R$. Thus
$\alpha_{\msbar}$ can be read directly off fig~\ref{falpha}. Since
the behaviour is consistent with two-loop perturbation theory, we
may express this in the conventional way in terms
of a $\Lambda$ parameter. We can try to make a best fit to this data,
however the smallest $R$ point is suspect since the lattice artifact
correction is presumably largest, and the
larger $R$ points are progressively
affected by non-perturbative components. Thus we prefer to quote
a band of values which encompass our results: $a \Lambda_R=0.041(3)$.
Expressed in the conventional way this gives $\sqrt K / \Lambda_L =
32.1(2.8)$.  This result is in agreement with that found by a similar
analysis~\cite{cmlms} at $\beta=2.7$ namely 31.9(1.7). That
agreement provides further justification for our procedure for removing
the lattice artifacts from the force at small $R$. Because
of the smaller lattice spacing in our present work, we are now
able to push to smaller $R$-values ($R \sqrt K = 0.1$ ). This gives
more confidence that the two loop perturbative expression for the
force is accurate and hence determines $\Lambda$ more reliably.

A different method of extracting the running coupling constant from
a study of finite volume effects on the lattice has
 recently been used by L\"uscher et al.~\cite{luco}.
 They analyse a succession of
small-volume lattices to track the behaviour of the running coupling
into the perturbative region. As a measurable they use the
response to a constant colour electric field. They are able to reach scales
corresponding to $R \sqrt K=0.05$ which is smaller than that we
can reach here with confidence.
They study the lattice size $L/a$ needed to have constant renormalised
coupling as $\beta$ varies from 2.5 to 3.0. From this they obtain
$a(\beta)$ and their result is in close agreement with asymptotic
(to 2 loops) scaling.  This result is in contrast with the large
volume analyses (see above) which find $a(\beta)$ to decrease
significantly faster
with increasing $\beta$. These two observations suggest that
the approach to asymptotic scaling  differs in small volumes from
 large ones.

If that is the case, then it is difficult for them to set the scale
 since the largest lattice size $L_{8}$
 used by~\cite{luco} corresponds to $z=m(0^{++})L_{8}
=3.2$. However, for $z < 5 $, the glueball and
string tension results are significantly different from those
on a large lattice~\cite{tickle}. The effect of adding a constant
colour electric field is unlikely to change this conclusion that
$z < 5 $ is very different from the large volume domain.
 Indeed the
simulations of ~\cite{luco} are in the $z$-region which is well described
by the semi-analytic methods of Van Baal ~\cite{vanb}. It would
be interesting to extend those calculations to include a constant
colour electric field and make a direct comparison with ref~\cite{luco}.
Indeed it might be expected that scaling close to
two-loop perturbative scaling would then be found since the
calculation leans heavily on perturbation theory and the (relatively)
trivial vacuum structure of small volumes.

\begin{table}
\begin{center}
\begin{tabular}{|r|l|c|l|}\hline
 $R/a$ \  & $\ \Delta V/\Delta R $  & $\Delta V_c / \Delta R$ &
 $ \qquad \alpha(R)$  \\\hline
 1.2071 & 0.1464(2)    & 0.1175    & 0.2216(3)(65)    \\
 1.5731 & 0.0826(4)    & 0.0750    & 0.2451(13)(25)   \\
 1.8660 & 0.0187(9)    & 0.0568    & 0.2623(40)(177)   \\
 2.1180 & 0.0686(4)    & 0.0463    & 0.2762(22)(131)  \\
 2.3428 & 0.0473(8)    & 0.0404    & 0.2948(57)(56)  \\
 2.6390 & 0.0306(5)    & 0.0326    & 0.3007(42)(20)  \\
 2.9142 & 0.0159(15)   & 0.0277    & 0.3136(172)(136)   \\
 3.0811 & 0.0172(19)   & 0.0247    & 0.3123(245)(106)  \\
 3.2395 & 0.0276(9)    & 0.0234    & 0.3279(126)(54)   \\
 3.3904 & 0.0267(22)   & 0.0245    & 0.3747(333)(70) \\
 3.5348 & 0.0165(28)   & 0.0198    & 0.3290(473)(72) \\
 3.6736 & 0.0212(19)   & 0.0200    & 0.3599(337)(28) \\
 3.9324 & 0.0180(5)    & 0.0188    & 0.3868(98)(5)  \\
 4.1231 & 0.0144(33)   & 0.0160    & 0.3720(779)(42) \\
 4.2426 & 0.0163(24)   & 0.0179    & 0.4425(591)(18) \\
 4.3589 & 0.0147(6)    & 0.0144    & 0.3917(163)(4)  \\
 4.6904 & 0.0129(4)    & 0.0128    & 0.4174(140)(2)  \\\hline
\end{tabular}
\caption{ The force $\Delta V/\Delta R$ and lattice artifact corrected force
$ \Delta V_c / \Delta R $ at
average separation $R$. The running coupling $\alpha (R)$ derived
from the corrected force is shown as well. The second error shown on
$\alpha$ is $10\%$ of the lattice artifact correction.}
\label{alpha}
\end{center}
\end{table}

\section{Conclusions}

We have explored the lattice simulation of SU(2) gauge theory with
the smallest lattice spacing used so far in a large volume study.
It is important to keep the volume sufficiently large to avoid
pronounced finite lattice effects. Indeed $\sqrt K L > 2 $ is needed
{}~\cite{tickle}. Here we have $\sqrt K L = 3$. To give a clearer
indication of our results we set the scale using $\sqrt K = 0.44 $GeV. Then
our lattice spacing $a=0.14$ GeV $=0.028 $ fm while the spatial lattice
size is 1.3 fm.

We find excellent scaling of large volume physical quantities for
$\beta \ge 2.4$. Thus the
force $dV(R)/dR$ is a universal function of $R$ over a range of
lattice spacings varying by a factor of 4. This, combined with evidence
from   glueball spectra, points
to an excellent agreement with scaling for all physical measurables
in pure gauge SU(2) simulation at large volume. Recent small volume
simulations~\cite{luco} suggest that the behaviour $a(\beta)$
may be different in that case for the $\beta-$range we have explored.
This would imply that the approach to asymptotic scaling is finite
size dependent.

We do not find agreement with two-loop perturbative asymptotic scaling.
 Indeed the effective $\beta$-function is only $80\%$ of the
perturbative
expression evaluated in terms of the bare lattice coupling constant from
$\beta=4/g^2$. This deviation can be ascribed to
the bad choice of the lattice bare coupling constant as an expansion
parameter. Using instead the small-$R$ force to define a renormalised
coupling constant, we find a {\it running} coupling which varies
as the two-loop perturbative expression. We explore this coupling
down to scales $R \sqrt K = 0.1 $ which corresponds to $R^{-1}= 4.4$
GeV. This is a sufficiently hard momentum that the perturbative
expression should be reliable and the scale $\Lambda$ can be estimated.
We find our running coupling results can be described by $\Lambda_{\msbar}
=272(24)$ MeV.

This work thus establishes that lattice gauge theory simulation
is a viable method to extract continuum results. Further decrease in
lattice spacing $a$ beyond that reported here is not expected to
lead to any change in our results for ratios of physical quantities.
 We have extended this approach to
SU(3) ~\cite{ukqcdsu3} in the quenched approximation but
the further extension to full QCD with dynamical fermions is still
needed and is a considerable computational challenge.

\bigskip
\noindent{\bf Acknowledgements}
\par
The computations using the Meiko Computing Surface at Edinburgh
were supported by SERC under  grants
GR/G 32779 and GR/H 49191, by the University of Edinburgh and by
 Meiko Ltd. The computations on the Meiko Computing surface at
Liverpool were supported by SERC under grant GR/G 37132 and EC
contract SC1 *CT91-0642.
The computations using the CRAY were supported by SERC grant GR/H 01236.
We acknowledge a useful correspondence with Martin L\"uscher.

\newpage

\begin{figure}[t]
\centering
\vspace{10cm}
\includegraphics{fig1.ps}
\caption{
The  force  between  static  quarks  versus  separation $R$   at
$\beta =2.85$.   The  force (see table~\protect\ref{tablefor})  is   defined
  as   a   finite   difference
$a[V(R+a)-V(R-a)]/2$. The fit to the force for $R\ge 6a$ is shown by the
 curve.
\label{fforce}
}
\end{figure}

\newpage

\begin{figure}[h]
\centering
\vspace{10cm}
\includegraphics{fig2.ps}
\caption{
The ratio between the potential values from table~\protect\ref{table3d} and
the fit with parameters given in table~\protect\ref{par3d}. As well as
this fit with a lattice one gluon exchange component, we show (open
triangles) the
result of not making this correction for lattice artifacts.
\label{offax}
}
\end{figure}

\begin{figure}[h]
\centering
\vspace{10cm}
\includegraphics{fig3.ps}
\caption{
The force at different $\beta$-values compared using $\alpha (R) =
{4 \over 3}R^2 \Delta V_c(R) / \Delta R$,
 where $V_c$ is the potential corrected
for lattice artifacts. Data at $\beta=2.85$ are from
table~\protect\ref{tablefor} (diamonds),
at $\beta=2.7$ (triangles) from ref~\protect\cite{cmlms} and
 at $\beta=2.4$ (squares) from ref~\protect\cite{cmtok}.
 Scaling corresponds to multiplying the $R$-axis by a
constant as described in the text.
\label{scaling}
}
\end{figure}


\begin{figure}[b]
\centering
\vspace{10 cm}
\includegraphics{fig4.ps}

\caption{
The effective running coupling constant from table~\protect\ref{alpha}
versus $R$ at $\beta=2.85$. The statistical (continuous) and
 systematic (dotted) errors are shown.
  The curves represent the evolution
expected from the two loop beta function with $a\Lambda_R = 0.044$
(dotted) and 0.038 (continuous).
\label{falpha}
}
\end{figure}


\newpage

\begin{thebibliography}{99}

\bibitem{ukqcd}
S.P.~Booth et al (the UKQCD Collaboration), Phys.\ Lett.\ B 275
(1992) 424.

\bibitem{deckandfor}
K.~Decker and  P.~de~Forcrand,  Nucl.\ Phys.\ B (Proc.  Suppl)
17(1990) 567.

\bibitem{kenandpen}
A.~Kennedy and B.~Pendleton, Phys.\ Lett.\ 156B (1985) 393.

\bibitem{ahranddiet}
J.~Ahrens and U.~Dieter, Math.\ J.\ of  Comp.\ 27 (1973) 927.

\bibitem{teper}
M.~Teper, Phys.\ Lett.\ 183B (1987) 345.

\bibitem{albanese}
M.~Albanese et al., Phys.\ Lett.\ 192B (1987) 163.

\bibitem{perandmich91}
S. J. Perantonis and C. Michael, Nucl.\ Phys.\ B(Proc. Suppl) 20
(1991) 177; J. Phys. G (in press).

\bibitem{perhuntandmich}
S.~J.~Perantonis, A.~Huntley and C.~Michael, Nucl.\ Phys.\ B326 (1989) 544

\bibitem{cmlms}
C.~Michael, Phys. Lett. 283B (1992) 103.

\bibitem{hellandkar}
U.~Heller and F.~Karsch, Nucl.\ Phys.\ B251 (1985) 254.

\bibitem{lm}
P. Lepage and P. Mackenzie, Nucl. Phys. B (Proc. Suppl.) 20 (1991) 173.

\bibitem{cmtok}
C. Michael, Nucl. Phys. B (Proc. Suppl) 26(1992) 417.

\bibitem{luco}
M. L\"uscher, R. Sommer, U. Wolff and P. Weisz , Cern preprint
TH 6566/92, DESY preprint 92-096.

\bibitem{tickle}
C. Michael, G. A. Tickle and M. Teper, Phys. Lett. B 207 (1988) 313.

\bibitem{vanb}
J. Koller and P. van Baal, Nucl. Phys. B302 (1988) 1.

\bibitem{ukqcdsu3}
UKQCD collaboration, in preparation.

\end{thebibliography}
\end{document}